\documentclass{llncs}
\usepackage{epsfig}

\usepackage{amssymb, amsmath} 
\usepackage{color}
\usepackage{times}
\usepackage{amsfonts}
\usepackage{graphicx,subfigure}
\usepackage{epsfig}
\usepackage[english]{babel}
\usepackage{hyperref}
\usepackage{verbatim}
\usepackage{xspace}

\title{Computational Topology Techniques for Characterizing Time-Series Data}

\author{Nicole Sanderson, Elliott Shugerman, Samantha Molnar, James
  D. Meiss, and Elizabeth Bradley}
 \institute{University of Colorado, Boulder CO USA}

\begin{document}
\maketitle

\newcommand{\todo}{\textcolor{blue}}
\newcommand{\discuss}{\textcolor{green}}
\newcommand{\chg}{\textcolor{red}}

\begin{abstract}

Topological data analysis (TDA), while abstract, allows a
characterization of time-series data obtained from nonlinear and
complex dynamical systems.  Though it is surprising that such an
abstract measure of structure---counting pieces and holes---could be
useful for real-world data, TDA lets us 
compare different systems, and even do membership testing or
change-point detection. However, TDA is computationally expensive and
involves a number of free parameters. This complexity can be obviated
by coarse-graining, using a construct called the witness
complex. The parametric dependence gives rise to the concept of
persistent homology: how shape changes with scale.  Its results allow
us to distinguish time-series data from different systems---e.g., the
same note played on different musical instruments.

\end{abstract}

\noindent {\bf Citation:} {N. Sanderson, E. Shugerman, S. Molnar, J. Meiss, and E. Bradley, ``Computational Topology Techniques for Characterizing Time-Series Data", {\sl IDA-17 (Proceedings of the 13th International Symposium on Intelligent Data Analysis)}, London, October 2017.

\medskip

\noindent {\bf Acknowledgment:} This material is based upon work sponsored by the National Science Foundation (award \#1245947). Any opinions, findings, and conclusions or recommendations expressed in this material are those of the author(s) and do not necessarily reflect the views of the NSF.

\section{Introduction}
\label{sec:intro}

Topology gives perhaps the roughest characterization of shape,
distinguishing sets that cannot be transformed into one another by
continuous maps \cite{Munkres84}.  The Betti numbers $\beta_k$, for
instance, count the number of $k$-dimensional ``holes" in a set:
$\beta_0$ is the number of components, $\beta_1$ the number of
one-dimensional holes, $\beta_2$ the number of trapped volumes, etc.
Of course, measures that are this abstract can miss much of what is
meant by ``structure,'' but topology's roughness can also be a virtue
in that it eliminates distinctions due to unimportant distortions.
This makes it potentially quite useful for the purposes of
classification, change-point detection, and other data-analysis
tasks\footnote{The work we describe in this paper calls upon areas of
  mathematics---including dynamical systems, topology and persistent
  homology---that may not be commonly used in the data-analysis
  community. As a full explanation of these would require several
  textbook length treatments, we content ourselves with discussing how
  these ideas can be applied, leaving the details of the theory to
  references.}.

Applying these ideas to real-world data is an interesting challenge:
how should one compute the number of holes in a set if one only has
samples of that set, for instance, let alone if those samples are
noisy?  The field of topological data analysis (TDA)
\cite{Kaczynski04,Zomorodian12} addresses these challenges by building
     {\sl simplicial complexes} from the data---filling in the gaps
     between the samples by adding line segments, triangular faces,
     etc.---and computing the ranks of the homology groups of those
     complexes.  These kinds of techniques, which we describe in more
     depth in \S\ref{sec:tda}, have been used to characterize and
     describe many kinds of data, ranging from molecular structure
     \cite{Xia2015} to sensor networks \cite{deSilva2007}.
     
As one would imagine, the computational cost of working with a
simplicial complex built from thousands or millions of data points can
be prohibitive.  In \S\ref{sec:tda} we describe one way, the {\sl
  witness complex}, to coarse-grain this procedure by downsampling the
data.  Surprisingly, one can obtain the correct topology of the
underlying set from such an approximation if the samples satisfy some
denseness constraints \cite{Alexander15}.  The success of this
coarse-graining procedure requires not only careful mathematics, but
also good choices for a number of free parameters---a challenge that
can be addressed using {\sl persistence}
\cite{Edelsbrunner02,Robins02}, an approach that is based on the
notion that any topological property of physical interest should be
(relatively) independent of parameter choices in the associated
algorithms.  This, too, is described in \S\ref{sec:tda}.

In this paper, we focus on time-series measurements from dynamical
systems, with the ultimate goal of detecting bifurcations in the
dynamics---change-point detection, in the parlance of other fields.
Pioneering work in this area was done by Muldoon {\sl et
  al.}~\cite{Muldoon93}, who computed Euler characteristics and Betti
numbers of embedded trajectories.  The scalar nature of many
time-series datasets poses another challenge here.  Though it is all
very well to think about computing the topology of a state-space
trajectory from samples of that trajectory, in experimental practice
it is rarely possible to measure every state variable of a dynamical
system; often, only a single quantity is measured, which may or may
not be a state variable---e.g., the trace in Fig.~\ref{fig:dce}(a), a
time series recorded from a piano.
\begin{figure}
\includegraphics[width=\textwidth, keepaspectratio]{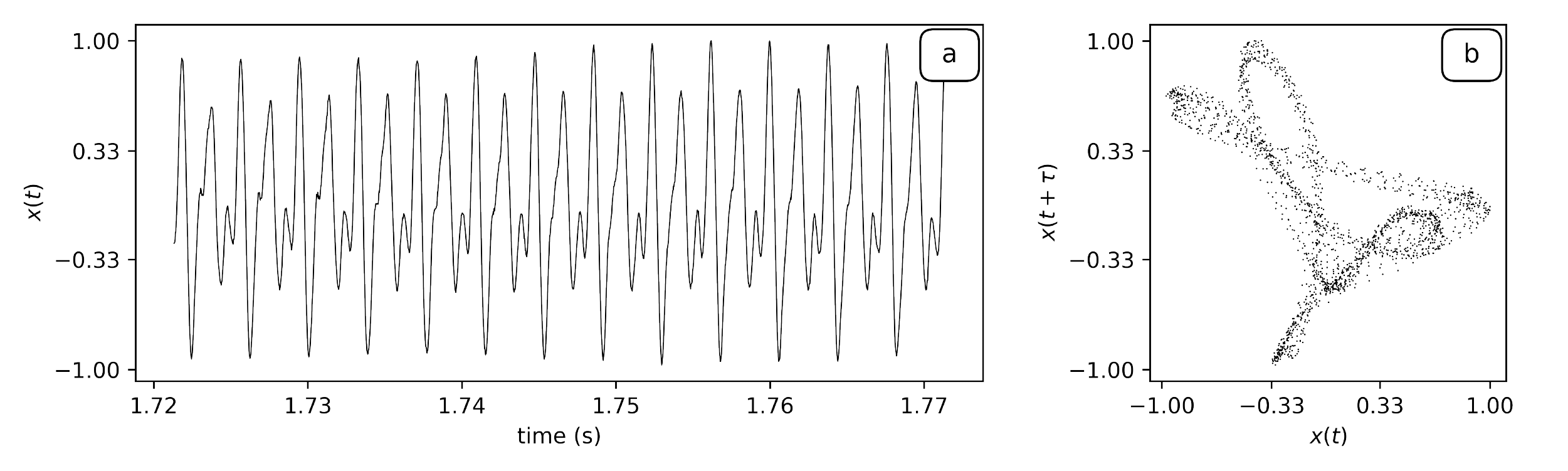}
    \caption{A short segment (45 msec) of a recording of middle C
      ($\mathfrak{f} = 261.62$ Hz) played on a Yamaha upright piano,
      recorded at 44100 Hz sample rate using a Sony ICD-PX312 digital
      voice recorder: (a) time series data (b) two-dimensional delay
      reconstruction using $\tau = \frac{1}{\mathfrak{f}\pi}$.}
    \label{fig:dce}
\end{figure}
The state space of this system is vast: vibration modes of every
string, the movement of the sounding board, etc. Though each quantity
is critical to the dynamics, we cannot hope to measure all of them.
Delay reconstruction \cite{BradleyKantz2015} lets one reassemble the
underlying dynamics---up to smooth coordinate change, ideally---from a
single stream of data.  The coordinates of each point in such a
reconstruction are a set of time $\tau$ delayed measurements $x(t)$:
from a discrete time series $\{x_t\}_{t=1}^N$, one constructs a
sequence of vectors $\{ \textbf{x}_t \}_{t = d_E \tau}^N$ where
$\textbf{x}_t = (x_t, x_{t-\tau}, \hdots, x_{t-(d_E-1)\tau})$ that
trace out a trajectory in a $d_E$-dimensional reconstruction space.
An example is shown in Fig.~\ref{fig:dce}(b).  Because the
reconstruction process preserves the topology---but not the {\sl
  geometry}---of the dynamics, a delay reconstruction can look very
different than the true dynamics.  Even so, this result
means that if we can compute the topology of the
reconstruction,
we can assert that the results hold for the underlying dynamics, whose
state variables we do not know and have not measured.  In other words,
the topology of a delay reconstruction can be useful in identifying
and distinguishing different systems, even if we only have incomplete
measurements of their state variables, and even though the
reconstructed dynamics do not have the same geometry as the originals.

Like the witness-complex methodology, delay reconstruction has free
parameters.  A reconstruction is only {\sl guaranteed} to have the
correct topology---that is, to be an ``embedding''---if the delay
$\tau$ and the dimension $d_E$ are chosen properly.  Since we are
using the topology as a distinguishing characteristic, that
correctness is potentially critical here.  There are theoretical
guidelines and constraints regarding both parameters, but they are not
useful in practice.  For real data and finite-precision arithmetic,
one must fall back on heuristics to estimate values for these
parameters \cite{fraser-swinney,KBA92}, a procedure that is subjective
and sometimes quite difficult.  However, it is possible to compute the
coarse-grained topology of 2D reconstructions like the one in
Fig.~\ref{fig:dce}(b) even though they are not true embeddings
\cite{GarlandBradleyMeiss16}.  This is a major advantage not only
because it sidesteps a difficult parameter estimation step, but also
because it reduces the computational complexity of all analyses that
one subsequently performs on the reconstruction.

This combination of ideas---a coarse-grained topological analysis of
an incomplete delay reconstruction of scalar time-series data---allows
us to identify, characterize, and compare dynamical systems
efficiently and correctly, as well as to distinguish different ones.
This advance can bring topology into the practice of data analysis, as
we demonstrate using real-world data from a number of musical
instruments.

\section{Topological data analysis}
\label{sec:tda}

There has been a great deal of work on change-point detection in data
streams, including a number of good papers in past IDA symposia (e.g.,
\cite{krishnan-IDA09}).  Most of the associated techniques---queueing
theory, decision trees, Bayesian techniques, information-theoretic
methods, clustering, regression, and Markov models and classifiers
(see, e.g., \cite{domingos06,gama10,keogh11,song05})---are based on
statistics, though frequency analysis can also play a useful role.
Though these approaches have the advantages of speed and noise
immunity, they also have some potential shortcomings.  If the regimes
are dynamically different but the operative distributions have the
same shapes, for example, these methods may not distinguish between
them.  They implicitly assume that it is safe to aggregate
information, which raises complex issues regarding the window size of
the calculation.  Most of these techniques also assume that the
underlying system is linear.  If the data come from a nonstationary
but deterministic nonlinear dynamical system---a common
situation---all of these techniques can fail.  Our premise is that
computational topology can be useful in such situations; the challenge
is that it can be quite expensive.

The foundation of TDA is the construction of a simplicial complex to
describe the underlying manifold of which the data are a (perhaps
noisy) sample: that is, to reconstruct the solid object of which the
points are samples.  A simplicial complex is, loosely speaking, a
triangulation.  The data points are the vertices,
edges---one-simplices---join those vertices, two-simplices cover the
faces, and so on. Abstractly, a $k$ simplex is an ordered list $\sigma
= \{x_1,x_2,\ldots, x_{k+1}\}$ of $k+1$ vertices.  The mathematical
challenge is to connect the data points in geometrically meaningful
ways.  Any such solution involves some choice of scale $\epsilon$: a
discrete set of points is only an approximate representation of a
continuous shape and is accurate only up to some spatial scale.  This
is both a problem and an advantage: one can glean useful information
from investigations of how the shape changes with $\epsilon$
\cite{ida04}.  While topology has many notions of shape, the most
amenable to computation is homology, which determines the Betti
numbers mentioned in \S\ref{sec:intro}. Computing these as a function
of $\epsilon$ is the fundamental idea of persistent homology, as
discussed further below.

There are many ways to build a complex. In a \v{C}ech complex, there
is an edge between two vertices if the two balls of radius
$\frac{\epsilon}{2}$ centered at the vertices intersect; here the
selection of $\epsilon$ fixes the scale.  Similarly, three vertices in
a \v{C}ech complex are linked by a two-simplex if the corresponding
three $\frac{\epsilon}{2}$-balls have a common intersection, and so
on.  Fig.~\ref{fig:complexes}(a) shows a \v{C}ech complex
constructed from the points in Fig.~\ref{fig:dce}(b).
\begin{figure}
\includegraphics[width=\textwidth, keepaspectratio]{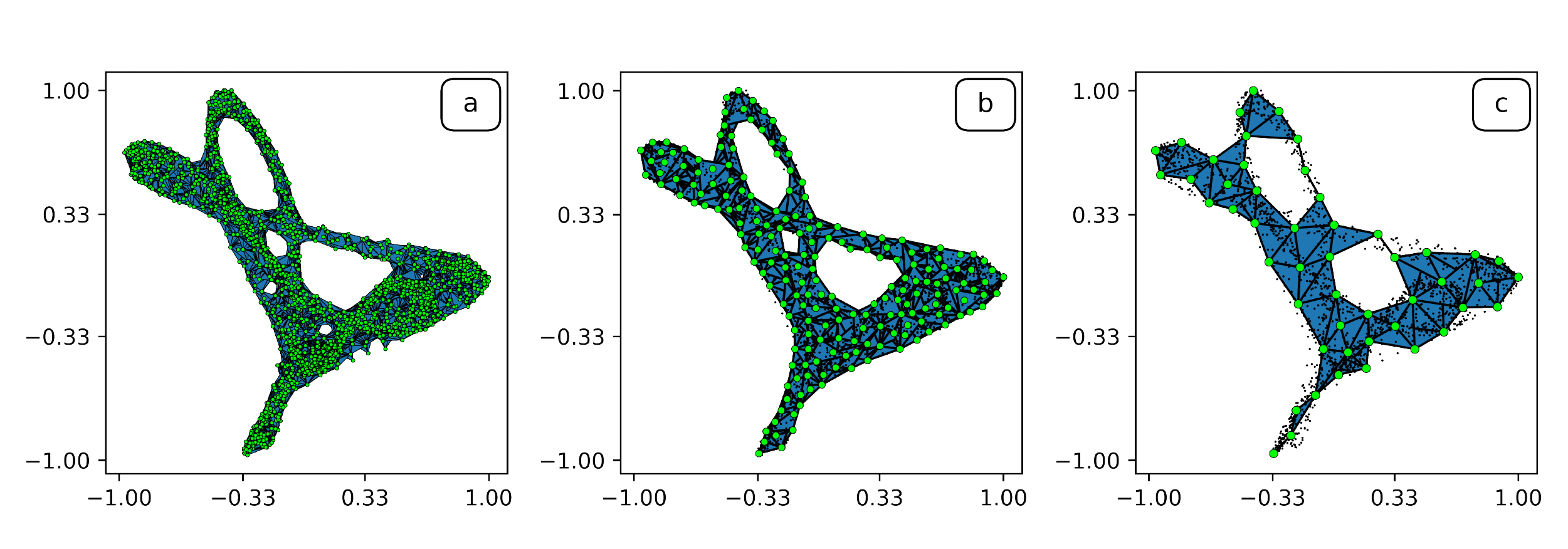}
\caption{Different simplicial complexes built from the data set in
  Fig.~\ref{fig:dce}(b) with $\epsilon=0.073$: (a) a \v{C}ech complex,
  with all $2000$ points used as vertices; (b) and (c) witness complexes
  with $\ell=200$ and $\ell=50$, respectively---i.e., with 1/10th and
  1/40th of the points used as landmarks.  Complexes (a) (b) and (c)
  contain $2770627$, $3938$, and $93$ triangles, respectively.}
\label{fig:complexes}
\end{figure}
The distance checks involved in building such a complex---between all
pairs of points, all triples, etc.---are computationally impractical
for large data sets.  There are many other ways to build simplicial
complexes, including the $\alpha$-complex \cite{Edelsbrunner94}, the
Vie\-tor\-is-Rips complex \cite{Ghrist08}, or even building a complex
based on a cubical grid \cite{Kaczynski04}.  All of these approaches
have major shortcomings for practical purposes: high computational
cost, poor accuracy, and/or inapplicability in more than two
dimensions.

An intriguing alternative is to coarse-grain the complex, employing a
subset of the data points as vertices and using the rest to 
how to fill in the gaps.  One way to do this is a {\sl witness
  complex} \cite{deSilva04}, which is determined by the time-series
data, $W$ (the witness set) and a smaller, associated set $L$---the
landmarks, which form the vertices of the complex.  Key elements of
this process are the selection of appropriate landmarks, typically a
subset of $W$, and a choice of a {\sl witness relation} $R(W,L)
\subset W \times L$, which determines how the simplices tile the
landmarks: a point $w \in W$ is a witness to an abstract simplex
$\sigma \in 2^L$ whenever $\{w\} \times \sigma \subset R(W,L)$.  One
connects two landmarks with an edge if they share at least one
witness---this is a one-simplex.  Similarly, if three landmarks have a
common witness, they form a two-simplex, and so on.  (This is similar
to the \v{C}ech complex, except that not every point is a vertex.)

There are many ways to define what it means to share a witness.
Informally, the rationale is that one wants to ``fill in'' the spaces
between the vertices in the complex if there is at least one witness
in the corresponding region.  Following this reasoning, we could
classify a witness $w_i \in W$ as shared by landmarks $l_j,l_k \in L$
if $-\epsilon < |l_j-w_i| - |l_k-w_i|<\epsilon$---that is, if it is
roughly equidistant to both of them---and add an edge to the complex
if we find such a witness.  If the set of witnesses included a point
that were shared between three landmarks, we would add a face to the
complex, and so on.  That particular definition is problematic,
however: it classifies an $\epsilon$-equidistant witness as shared
even if it is on the opposite side of the data set from the two
landmarks.  To address this, we add a distance constraint to the
witness relation, classifying a witness $w_i \in W$ as shared between
two landmarks $l_j,l_k \in L$ if both are within $\epsilon$ of being
the closest landmark to $w_i$, i.e., if
	$\max(|l_j-w_i|, |l_k-w_i|) < \min_{m} |l_m-w_i| + \epsilon$:
\smallskip

\underline{Input:}  $\{x_t\}_{t = 1}^N$, discrete $\mathbb{R}$-valued time series

 \textbf{delay coordinate reconstruction}, $\{\textbf{x}_t\}_{t = d_E\tau}^N$
\

select landmarks $L = \{l_i\}_{i=1}^{\ell} \subseteq$ \{\textbf{x}$_t\}_{t = d_E\tau}^N$
\

compute pair-wise distances $D_{ij} = \vert l_i - \textbf{x}_j \vert $
\

 \textbf{for} $\epsilon \in (\epsilon_{\text{min}}, \epsilon_{\text{max}}, \epsilon_{\text{step}}):$ (build witness complex, $\mathcal{W}^{\epsilon}$)
\

\hspace{.5 cm}  \textbf{for} $\textbf{x}_t \in X$: 
  \
 
\hspace{1 cm} $\text{d} = \vert L - \textbf{x}_t \vert+ \epsilon$
\

 \hspace{1 cm} \textbf{for} $(l_i, l_j) \in L:$  \hspace{.2cm} (check for edges)
\

 \hspace{1.5 cm} \textbf{if} $\vert l_i - \textbf{x}_t \vert, \vert l_j - $\textbf{x}$_t \vert < \text{d}:$
  	\
	
	\hspace{2. cm}$\{l_i, l_j\} \in \mathcal{W}^{\epsilon}$
\

 \hspace{1 cm} \textbf{for} $(l_i, l_j, l_k) \in L:$ \hspace{.2cm} (check for triangles)
  \
  
 \hspace{1.5 cm} \textbf{if} $\vert l_i - \textbf{x}_t\vert, \vert l_j - \textbf{x}_t \vert , \vert l_k - \textbf{x}_t \vert< \text{d}:$
  \
  
  	\hspace{2 cm}$\{l_i, l_j, l_k\} \in \mathcal{W}^{\epsilon}$
	\
	\vspace{-.5cm}
	\begin{center}
	\hspace{1 cm} \vdots  \hspace{.5 cm} (check for higher dimensional simplices)
	\end{center}
	\vspace{-.6cm}
	\
	
\underline{Output:} $\{\mathcal{W}^{\epsilon}\}_{\epsilon_{\text{min}}}^{\epsilon_{\text{max}}}$, series of witness complexes for specified $\epsilon$ range

	\vspace{2mm}

\noindent Fig.~\ref{fig:complexes}(b) shows a witness complex
constructed in this manner from the data of
Fig.~\ref{fig:complexes}(a), with one-tenth of the points chosen as
landmarks. 
The computation involved is much faster---an order of magnitude less
than required for Fig.~\ref{fig:complexes}(a).  
The scale factor $\epsilon$ and the number $\ell$ of landmarks have
critical implications for the correctness and complexity of this
approach, as discussed further below\footnote{Landmark choice is
  another issue.  There are a number of ways to do this; here, we
  evenly space the landmarks across the data.}.

Every simplicial complex has an associated set of homology groups,
which depend upon the structure of the underlying manifold: whether or
not it is connected, how many holes it has, etc.  This is a
potentially useful way to characterize and distinguish different
regimes in data streams.  An advantage of homology over homotopy or
some other more complete topological theory is that it can be reduced
to linear algebra \cite{Munkres84}.  Algorithms to compute homology
depend on computing the null space and range of matrices that map
simplices to their boundaries \cite{Kaczynski04}.
The computational complexity of these algorithms scales badly,
though---both with the number of vertices in the complex and with the
dimension of the underlying manifold.  In view of this, the
parsimonious nature of the witness complex is a major advantage.
However, an overly parsimonious complex, or one that contains spurious
simplices, may not capture the structure correctly.

The parsimony tradeoff plays out in the choices of both of the free
parameters in this method.  Fig.~\ref{fig:complexes}(b) and (c)
demonstrate the effects of changing the number of vertices $\ell$ in
the witness complex.  With 200 vertices, the complex effectively
captures the three largest holes in the delay reconstruction; if
$\ell$ is lowered to 50, the complex is too coarse to capture the
smallest of these holes.
In general, increasing $\ell$ will improve the match of the
complex to the data, but it will also increase the computational
effort required to build and work with that structure.
Reference \cite{GarlandBradleyMeiss16} explores the accuracy end of
this tradeoff; the computational complexity angle is covered in the
later sections of this paper.
\begin{figure}[h!]
\includegraphics[width=\textwidth, keepaspectratio]{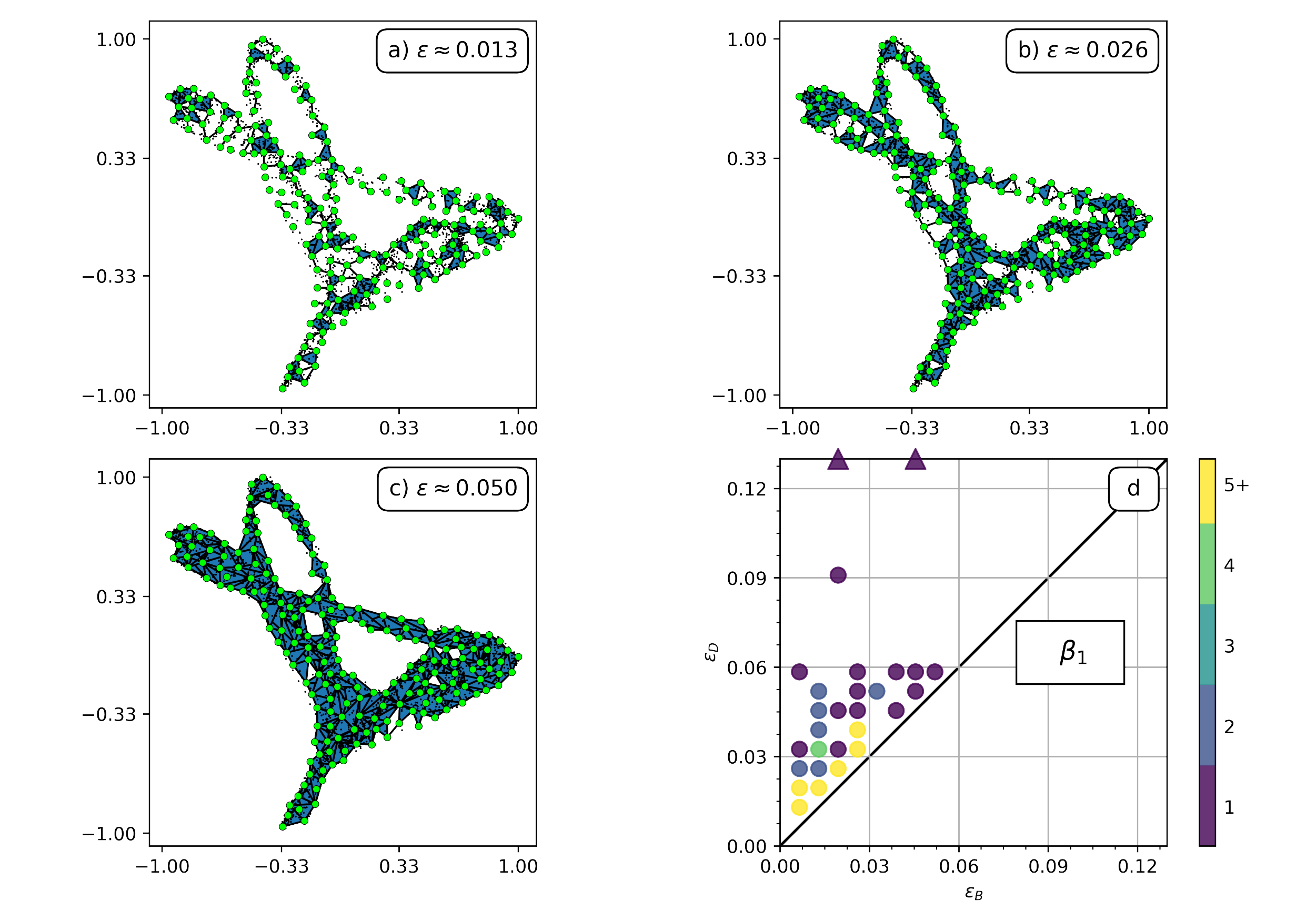}
\caption{The effects of the scale parameter $\epsilon$: (a)-(c) show
  witness complexes with $\ell=200$ and different $\epsilon$ values.
  (d) shows a $\beta_1$ persistence diagram computed across a range of
  $\epsilon$ values.
Each point in (d) represents a hole in the complex; its $x$ and $y$
coordinates show the $\epsilon$ values at which that hole appears and
disappears, respectively.  Holes that persist beyond the upper
$\epsilon_{\text{max}} \approx 0.13$ are shown with triangles.}
\label{fig:persistence}
\end{figure}

The other free parameter in the process, the scale factor $\epsilon$,
plays a subtler and more interesting role.  When $\epsilon$ is very
small, as in Fig.~\ref{fig:persistence}(a), very few witnesses are
shared and the complex is very sparse.  As $\epsilon$ grows, more and
more witnesses fall into the broadening regions that qualify them as
shared, so more simplices appear in the complex, fleshing out the
structure of the sampled manifold.  There is a limit to this, however.
When $\epsilon$ approaches the diameter of the point cloud, the
witness complex will be fully connected, which obscures the native
structure of the sampled set; well before that, simplices appear that
do not reflect the true structure of the data.  One effective way to
track all of this is the {\sl $\beta_1$ persistence diagram} of
\cite{Edelsbrunner02}, which plots the $\epsilon$ value at which each
hole appears in the complex, $\epsilon_B$, on the horizontal axis and
the value $\epsilon_D$ at which it disappears from the complex on the
vertical axis\footnote{Choosing the range and increment for $\epsilon$
  in such a plot requires some experimentation; in this paper, we use
  $\epsilon_{\text{step}} = 20$ and $\epsilon_{\text{max}}$ set for
  each instrument when the first $20$-dimensional simplex is
  witnessed.  This is a good compromise between effectiveness and
  efficiency for the data sets that we studied.}.  A persistence
diagram for the piano data, for example---part (d) of
Fig.~\ref{fig:persistence}---shows a cluster of holes that are born
and die before $\epsilon = 0.06$.  These represent small voids in the
data.  The three points near the top left of
Fig.~\ref{fig:persistence}(d) represent holes that are highly
persistent.

\vspace*{-2mm}
\section{Persistent Homology and Membership Testing}
\label{sec:results}
\vspace*{-1mm}

Cycles are critical elements of the dynamical structure of many
systems, and thus useful in distinguishing one system from another.  A
chaotic attractor, for example, is typically densely covered by
unstable periodic orbits, and those orbits provide a formal
``signature'' of the corresponding system \cite{cvitanovic88}.
Topologically, a cycle is simply a hole, of any shape or size, in the
state-space trajectory of the system.  The persistent homology methods
described above include some aspects of geometry, though, which makes
the relationship between holes and cycles not completely simple.
Musical instruments are an appealing testbed for exploring these
issues.
Of course, one can study the harmonic structure of a note from an
instrument, or any other time series, using frequency analysis or
wavelet transforms.  Because delay reconstruction transforms time into
space, it not only reveals which frequencies are present at which
points in the signal, as well as their amplitudes.  These
reconstructions also bring out subtler features; any deviation from
purely elliptical shape, for instance---or the kind of ``winding''
that appears on Fig.~\ref{fig:dce}(b)---signals the presence of
another signal and also gives some indication of its amplitude and
relative frequency.

Topological data analysis brings out those kinds of features quite
naturally.  The structure of the persistence diagrams for the same
note played on two different musical instruments, for instance, is
radically different, as shown in Fig.~\ref{fig:clarinet-and-viol}.
The witness complex of a clarinet playing the A above middle C
contains seven holes for $\epsilon < 0.05$.  Six of these holes die
before $\epsilon = 0.06$; they are represented by the points in the
lower left corner of the persistence diagram.  The other hole in the
complex triangulates the large loop in the center of the
reconstruction of Fig.~\ref{fig:clarinet-and-viol}(a).  This hole,
which remains open until the end of the $\epsilon$ range of the
calculation, is represented by the triangle in the top left corner of
the persistence diagram in Fig.~\ref{fig:clarinet-and-viol}(c).  The
viol reconstruction in Fig.~\ref{fig:clarinet-and-viol}(b), on the
other hand, contains over twelve holes that are born at low $\epsilon$
values, including many short-lived features depicted in the lower left
of the persistence diagram.  By $\epsilon = 0.10$, only four holes
remain open.  The smallest of these four features, which closes up
around $\epsilon = 0.23$, is represented by the point in the top left
of Fig. 4(d).  The three other holes, which remain open to the end of
the $\epsilon$ range of the calculation, are represented by the
colored triangles at the top left of
Fig.~\ref{fig:clarinet-and-viol}(d).

\begin{figure}
\includegraphics[width=\textwidth, keepaspectratio]{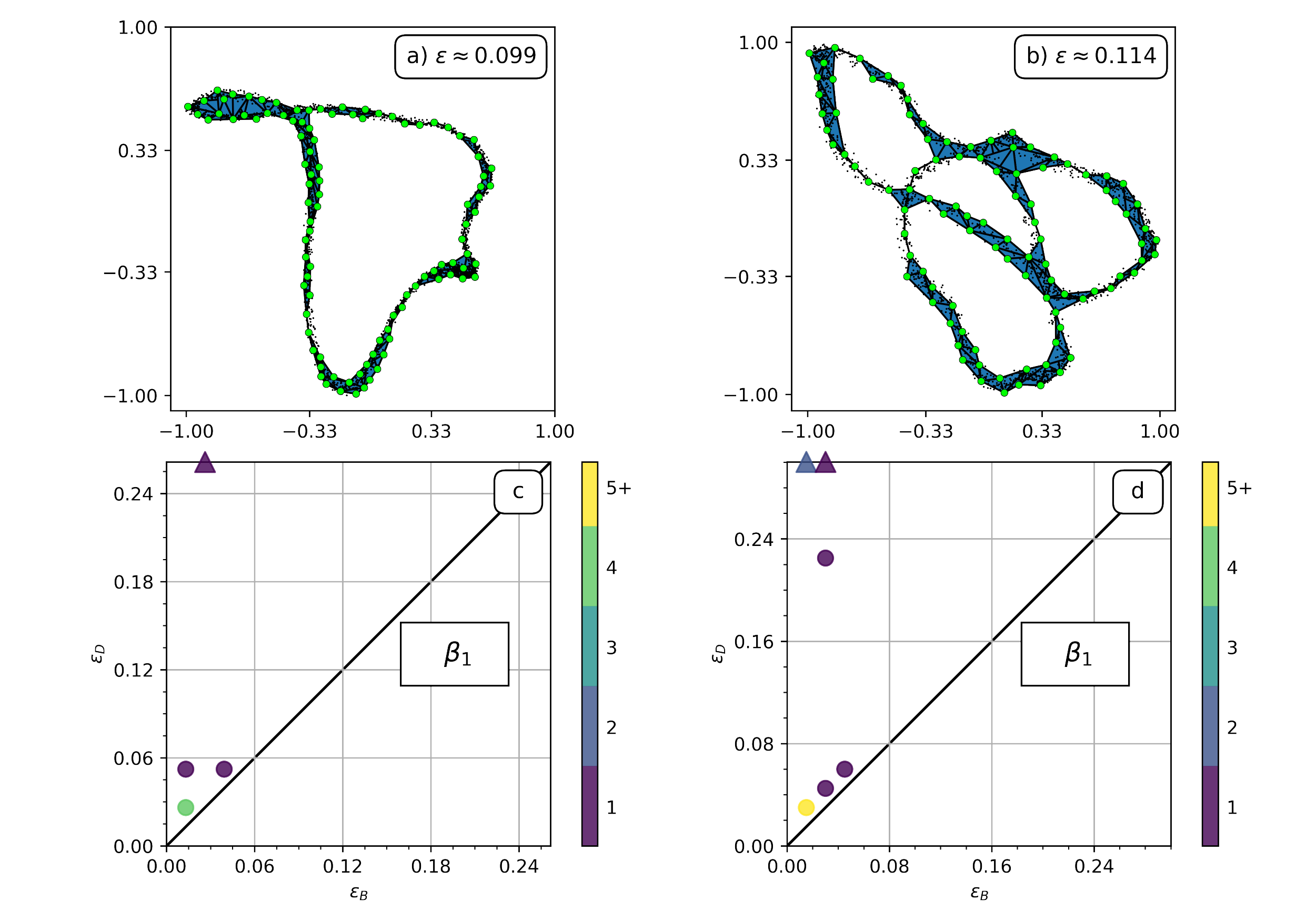}
\caption{Witness complexes for A440 ($\mathfrak{f} = 440$ Hz) played
  on a (a) clarinet and (b) viol constructed using $2000$ witnesses
  and $100$ landmarks for $\epsilon = 0.099$, $0.114 $ respectively.
  Beneath each are the corresponding $\beta_1$ persistence diagrams.
  The delay reconstructions are for approximately $0.05$ seconds each,
  with $\tau = \frac{1}{\mathfrak{f} \pi}$ seconds.}
\label{fig:clarinet-and-viol}
\end{figure}

The patterns in these persistence diagrams---the number of highly
persistent holes and short-lived features, and the $\epsilon$ values
at which they appear and disappear---suggest that computational
topology can be an effective way to distinguish between musical
instruments.  To test this more broadly, we built a pair of simple
classifiers that work with {\sl persistent rank functions} (PRFs),
cumulative functions on $\mathbb{R}^{2+}$ that report the number and
location of points in a persistence diagram \cite{Robins15}.  We
trained each classifier on 25 disjoint 0.05 sec windows from
recordings of the corresponding instrument.  This involved computing
the persistent homology for each instance, then computing the mean,
$\overline{\beta^1}$, and standard deviation, $\sigma$ of the set of
corresponding PRFs.  The test set comprised 50 0.05 sec windows, 25
from each instrument; for each of these samples, we computed the $L^2$
distance between the PRF of the sample and the mean
$\overline{\beta^1}$ for each instrument.  If that distance was below
$k \sigma$ for some threshold parameter $k$, we assigned membership in
the corresponding instrument class.  The receiver operating
characteristic (ROC) curves in the top row of Fig.~\ref{fig:ROCs}
plot the true positive rates versus the false positive rates for the PRF
classifiers.  The clarinet classifier achieves a true positive
rate 70\% around $k = 0.5$, and 100\% when $k = 1$.  The false
positive rate remains near 0\% up through $k = 5$, demonstrating a
broad range of threshold values $k$ for which the PRF classifier will
successfully assign membership in the clarinet class to most clarinet tones---and
non-membership to most viol tones.  The viol classifier
achieves a true positive rate near 70\% by $k = 1$ and over 90\% by $k
= 2$, maintaining a false positive rate below 50\% for all $k$ up to
$2.5$.

\begin{figure}
\includegraphics[width=\textwidth, keepaspectratio]{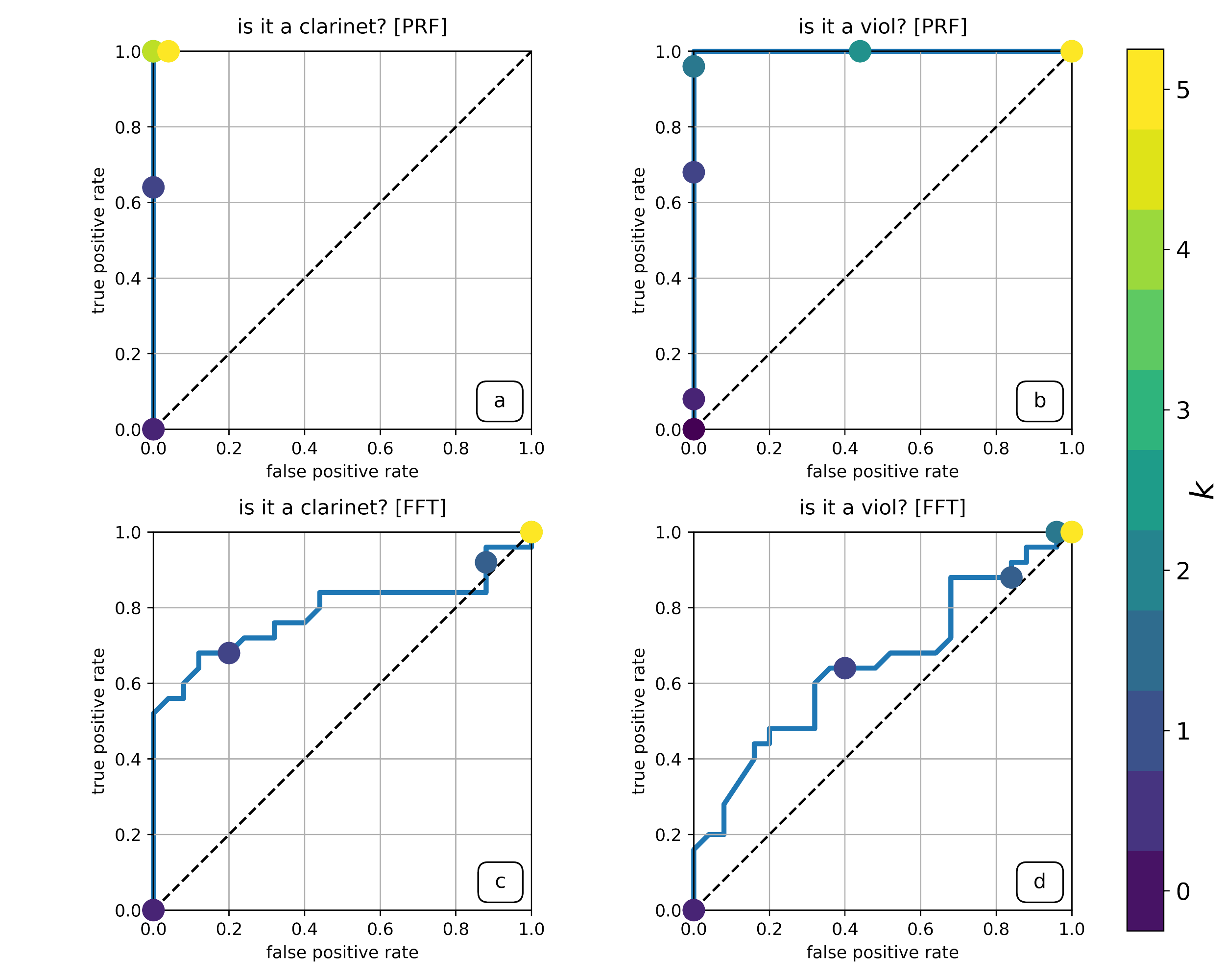}
\caption{ROC curves for a persistent homology-based classifier (top)
  and an FFT-based classifier (bottom) for clarinet (left) and viol
  (right) membership testing.  Color bar indicates the threshold
  parameter value $0<k<5$.}
\label{fig:ROCs}
\end{figure}

As a comparison, we built a pair of FFT-based classifiers, whose
results are shown in the bottom row of Fig.~\ref{fig:ROCs}, training
and testing them on the same samples used for the PRF-based
classifiers.  The feature vector in this case was a set of 2000
logarithmically spaced values between 10 Hz and 10,000 Hz from the
power spectrum of the signal.  As in the PRF-based classifier, we
computed the mean and standard deviation of this set of feature
vectors, classifying a sample as a viol or clarinet if its $L^2$
distance to the corresponding mean feature vector was less than $k
\sigma$.
As is clear from the shapes of the ROC curves, the PRF-based
classifiers outperformed the FFT-based classifiers.  The FFT-based
clarinet classifier achieves 70\% and 20\% true and false positive
rates, respectively, around $k = 0.5$.  Above that threshold, the
false positive rate rapidly catches up to the true positive rate,
making the classifier equally likely to correctly classify a clarinet
as a clarinet as it is to erroneously classify a viol as a clarinet.
The ROC curve for the FFT-based viol membership classifier is even
closer to the diagonal: it will correctly classify a viol as a viol
only slightly more often than it will erroneously classify a clarinet
as a viol, for any parameter value $0<k<5$.

Clarinets and viols produce very different sounds, of course, so
distinguishing between them is not a hugely challenging task.  A more
interesting challenge is to compare two pianos.  As shown in
Fig.~\ref{fig:two-pianos}, persistence diagrams of the same note
played on an upright piano and a grand piano are notably different:
the former has a single long-lived hole---the fundamental tone of the
note---while the latter has {\sl two}, perhaps reflecting the greater
sonic richness of the instrument.  Persistence diagrams for both
pianos contain many short-lived holes at low $\epsilon$ values, which
also speaks to a notable variance in the volumes of the resonating
frequencies.
\begin{figure}
\includegraphics[width=\textwidth, keepaspectratio]{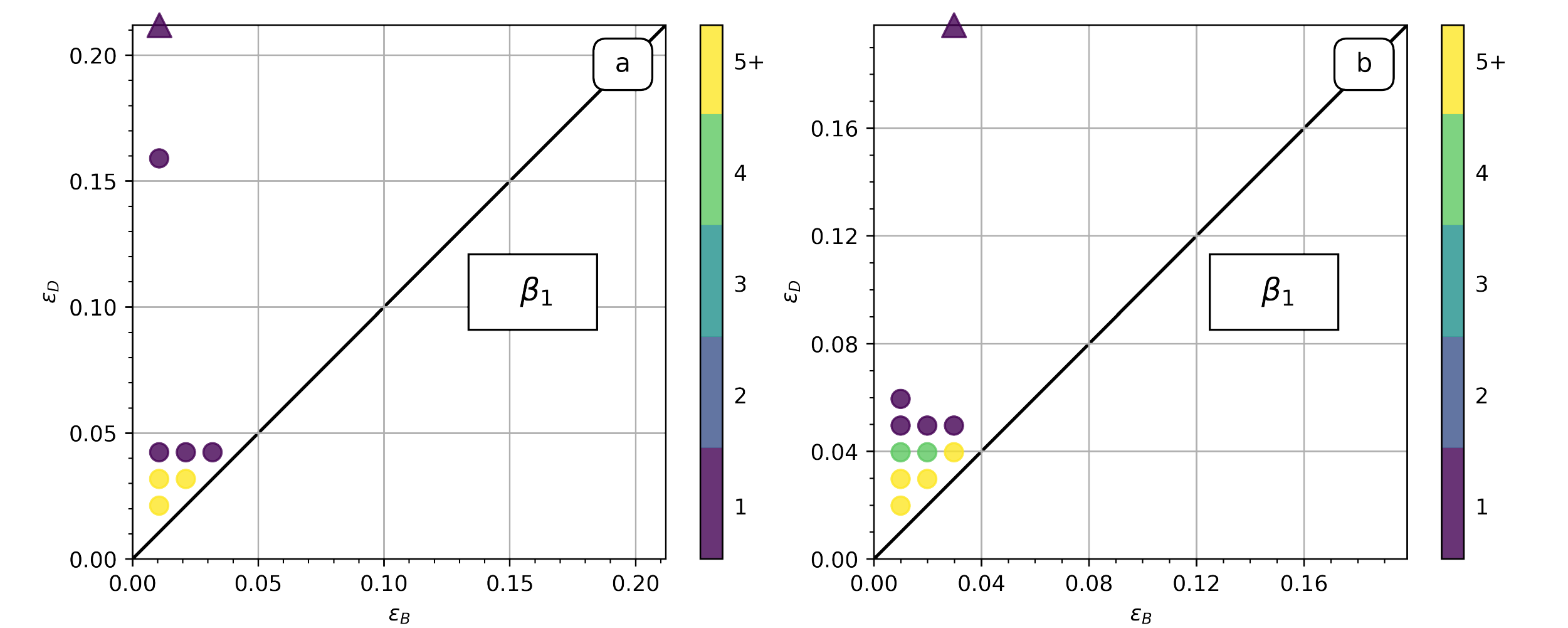}
\caption{Persistence diagrams for A440 on two different pianos.  (a):
  a Steinway grand piano; (b): a Baldwin
   upright piano.}
\label{fig:two-pianos}
\end{figure}

Table~\ref{tab:complexity} shows the runtime and memory costs involved
in the construction of some of the complexes mentioned in this paper.
\begin{table}
\begin{center}
\begin{tabular}{|c|c|c|c|}
\hline Number of landmarks & Runtime (sec) & Number of two-simplices &
Memory usage \\
\hline 2000 (\v{C}ech) & 59.1 & 2,770,627 & 9.3 MB \\
200 & 3.9 & 3,938 & 0.9 MB \\
50 & 0.8 & 93 & $<$ 0.1 MB\\ \hline
\end{tabular}
\end{center}
\caption{The computational and memory costs involved in constructing
  different simplicial complexes from the 2000-point reconstruction of
  Fig.~\ref{fig:dce}(b) with $\epsilon=0.073$ on an Ubuntu Linux
  machine with an Intel Core i5 1.70GHz × 4 CPU and 12GB of memory.}
\label{tab:complexity}
\end{table}
These numbers make it quite clear why the parsimonious nature of the
witness complex is so useful: using {\sl all} of the points as
landmarks is computationally prohibitive.  And that parsimony,
surprisingly, does not come at the expense of accuracy, as long as the
samples satisfy some denseness constraints \cite{Alexander15}.
Nonetheless, this is still a lot of computational effort; the
membership test process described here involves building the complex,
computing the homology, repeating those calculations across a range of
$\epsilon$ values, and perhaps computing a persistent rank function
from the results.  The associated runtime and memory costs will worsen
with increasing $\epsilon$, and with the size of the data set, so
computational topology is not the first choice technique for every IDA
application.  However, it can work when statistical- and
frequency-based techniques do not.

\vspace*{-2mm}
\section{Conclusion}
\label{sec:conclusion}
\vspace*{-1mm}

We have shown that persistent homology can successfully distinguish
musical instruments using witness complexes built from two-dimensional
delay reconstructions for a single note.  This approach does
not rely on the linearity or data aggregation of 
many traditional membership-testing techniques; moreover, topological data
analysis can outperform these traditional methods. Though the associated
computations are not cheap, the reduction in model order and the
parsimony of the witness complex greatly reduce the
associated computational costs.

Persistent homology calculations---on any type of simplicial
complex---work by blending geometry into topology via a scale
parameter $\epsilon$.  Their leverage derives from the patterns that
one observes upon varying $\epsilon$, which are presented here in the
form of persistence diagrams.  The witness complex uses
the scale parameter to obtain its natural parsimony.  While the
specific form of the witness relation used in this paper is a good
start, it can still create holes where none ``should'' exist, and vice
versa.  Better witness relations---factoring in the temporal ordering
and/or the forward images of the witnesses, or the curvature of those
paths---will be needed to address those issues.  This is particularly
important in the context of the kinds of reduced-order models that we
use here to further control the computational complexity.  An
incomplete delay reconstruction is a projection of a high-dimensional
structure onto a lower-dimensional manifold: an action that can
collapse holes, or create false ones.  Changing $\tau$ also alters the
geometry of a delay reconstruction.  Understanding the interplay of
geometry and topology in an incomplete embedding, and the way in which
the witness relation exposes that structure, will be key to bringing
topological data analysis into the practice of intelligent data
analysis.

\bibliographystyle{splncs03} \bibliography{refs}

\end{document}